\documentstyle[sprocl,psfig]{article}

\def\be{\begin{equation}}
\def\ee{\end{equation}}
\def\bea{\begin{eqnarray}}
\def\eea{\end{eqnarray}}
\def\mg{\big <}
\def\md{\big >}
\def\d{{\rm d}}

\def\hkpc{$h^{-1}$kpc }
\def\hMpc{$h^{-1}$Mpc }
\def\h3Mpc{h^{-3}{\rm Mpc}^3 }

\def\h3Mpcinv{h^{3}{\rm Mpc}^{-3} }

\begin{document}
\pagestyle{empty}

\title{DARK MATTER AND  GRAVITATIONAL LENSING}

\author{ Y. MELLIER }

\address{Institut d'Astrophysique de Paris, 98 bis boulevard Arago\\ 75014
Paris, France (mellier@iap.fr) \\
Observatoire de Paris, DEMIRM, 61 avenue de l'Observatoire,\\
75014 Paris, France}

\author{F. BERNARDEAU}

\address{Service de Physique Th\'eorique, CE Saclay \\
91191 Gif-sur-Yvette Cedex, France (fbernardeau@cea.fr)}

\author{L. VAN WAERBEKE}

\address{Max-Planck-Institut f\"ur Astrophysik, Karl Schwarzschild-Str.
1\\
85740 Garching, Germany (waerbeke@mpa-garching.mpg.de)}


\maketitle\abstracts{The last decade has  shown a considerable development of 
gravitational lensing for cosmology because it    
 probes the amount and the nature of dark matter, and provides 
  information on the density parameter $\Omega$, the cosmological
constant $\Lambda$ and the Hubble constant $H_{o}$. Therefore, 
 gravitational lensing 
 can  constrain the
cosmological scenario which gave birth to the Universe as it appears today. 
 The ongoing programs
and future projects  which are developing now all over the world show
that gravitational lensing is considered  as a major cosmological
tool for the coming  years as well. In this review, we summarize
some of the most recent advances in the fields relevant for the dark
matter issue. We will focus on the microlensing,
 the arc(let)s and the weak lensing studies. The  
possibility  to  check  the existence of a non-zero $\Lambda$  is
presented  elsewhere (see Fort et al. contribution).  } 
  
\section{Introduction}\label{sec:intro}
The present-day structuration of the Universe 
 likely formed from gravitational condensations of primordial fluctuations. 
 In a homogeneous and isotropic universe, the growth and the late
evolution of these fluctuations depend on the amount of mass-energy
presents in the Universe and the nature of its matter content. The former
is described by the cosmological parameters $\Omega$ and $\Lambda$,
whereas its nature can be inferred from the shape of the 
power spectrum of the initial
fluctuations and the amount of baryonic matter we can observe today. 
These crucial quantities are then among the most
challenging observing targets for the end of this century, and    
  motivated also the launches of MAP and Planck-Surveyor by the
beginning of 2000.
\\
The large variety of observational techniques applied over a wide
range of dynamical systems shows compelling evidence that 
 most of them are dominated by invisible matter.
Furthermore, it seems that the dark matter fraction increases with the
mass range of the systems. The amount of dark matter   
deduced from these studies leads to the conclusions that 
 \ \ (1)  dark matter  is the main component of the Universe,
 \ \ (2) the visible mass does not fully account for the baryonic mass permitted
from the  theoretical expectations of the Big Bang Nucleosynthesis (BBN), 
 allowing part of the dark matter to be baryonic,
 \ \ (3)  on the other hand, if the mass-to-light ratios inferred from 
observations are correct, dark matter cannot be only baryonic. \\
There is still room for controversy on    
these conclusions because the measurements of the 
 amount and the distribution of the matter  are
 {\sl indirect}:  \ \  (1)  2- and 3-dimension galaxy  surveys only depict 
the distribution of {\sl light};  \ \ (2)  the mass  of
gravitational systems are not simply inferred  because     
assumptions on the geometry of their mass and light distribution
profiles and on their dynamical stage are necessary. Some of these hypotheses 
are
much debated.  \ \ (3) Finally, the dynamical studies   
of large-scale galaxy flows which map the large-scale mass
distribution of the Universe are not yet conclusive because of
 the poorness of the  catalogs. \\
In fact,
the rotation curves of spiral galaxies seem to give by far the most robust 
estimates of the total mass of galaxies.     
Athough the mass-to-light ratios inferred for these galaxies requires
 that their
halos are dominated by dark matter, the amount needed is  compatible 
with the upper limit of the baryon fraction deduced from BBN, and does  
not require that  halos have non-baryonic content.
  Hence, the search for the
nature of dark matter in galaxies as well as for new robust mass estimators 
 is important.\\
Gravitational lensing effects  can directly probe    
deflecting masses  and can determine
without ambiguity the amount of matter present along the line-of-sight.
 Its astrophysical interest only raised after the
discoveries of the first multiply imaged quasar \cite{wcw},
the  gravitational arcs \cite{sfmp}$^{\!,\,}$ \cite{smfmc}$^{\!,\,}$ \cite{lp}
 and the arclets \cite{fpmms}.  But the
 on-going massive monitoring of microlensing events and the development of
large programs for mapping the large-scale structures by using weak
lensing make gravitational lensing effect  one of the most promising
tools to address some cosmological issues of the next ten years.
This review only focus on the recent results relevant 
 to dark matter issue. Section \ref{sec:lensequa} summarizes the
fundamental concepts and equations of gravitational lensing. In section \ref{sec:baryongal}, 
the latest microlensing experiments are presented. 
Sections \ref{sec:arclusters} and \ref{sec:wlcluster} are
devoted to investigations of clusters of galaxies and section
\ref{sec:wllss} to the promising investigation of large-scale
structures. 
\section{The equations of lensing}\label{sec:lensequa}
\begin{figure}
\centerline{\psfig{figure={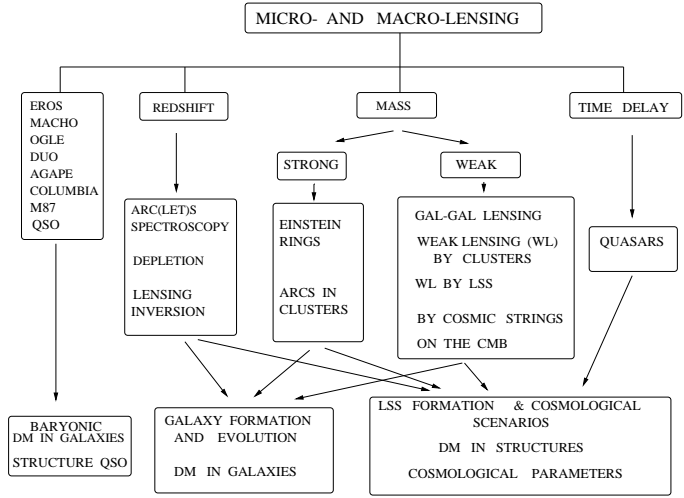},width=9. truecm}}
\caption{\label{application}Some applications of
gravitational lensing effects in cosmology. The left-hand box summarizes
the microlensing project. COLUMBIA is the pixel monitoring project of M31
done at KPNO and VATT. M87 is a similar project on M87 done with the
HST.}
\end{figure}

In presence of a gravitational field, 
 due  to the local modification of the geodesics, a background source 
located at 
the apparent position, $\vec \theta_S$, appears on the sky at the new apparent 
position $\vec \theta_I$ in presence of a gravitational field. The
lensing equation relates these apparent position to the deflection angle
$\vec \alpha$:
\be
\vec \theta_S=\vec \theta_I-\vec \alpha(\vec \theta_I)=\vec \theta_I-\vec \nabla_{\vec \theta_I}\varphi=\vec \theta_I-{2 \over c^2} {D_{ls} D_{ol} \over
D_{os}} \int \vec \nabla_{\vec \theta_I} \Phi(\vec \theta_I) dz \ ,
\label{eq:deviation}
\ee
where $\varphi$ is the projected lensing potential, $\Phi$ the 3-dimension
 newtonian potential, and the $D_{ij}$ are the angular distances with respect to
the observer $(o)$, the source $(s)$ and the lens $(l)$. The $D_{ij}$ 
express part of the dependence of the lensing equation with the cosmological
parameters (another dependence with $\Omega$ comes from the potential
$\Phi$ of the Poisson equation). The deformation
of a beam is given by the Jacobian of the mapping between the source 
 and the image planes, namely the amplification matrix $A$, whose terms are
\be
A_{ij}=(\delta_{ij}-{\partial \alpha_i(\theta) \over \partial
\theta_j})=(\delta_{ij}-{\partial^2 \varphi(\theta) \over
\partial\theta_i\partial\theta_j}) \ .
\label{eq:amplification1}
\ee
$A$ is usually expressed with the convergence $\kappa$, the isotropic 
term of the magnification, and the
gravitational shear $\gamma=\gamma_1+i \ \gamma_2$, the anisotropic term:
\be 
A=\pmatrix{ 1-\kappa-\gamma_1 & -\gamma_2 \cr
-\gamma_2 & 1-\kappa+\gamma_1 \cr } \ ,
\label{eq:amplification2}
\ee
where
\be
\kappa={1\over 2} \nabla^2 \varphi={\Sigma\over \Sigma_{\rm crit.}}; \ \ \
\gamma_
1={1\over 2}(\varphi_{,11}-\varphi_{,22}) \ ; \ \ \ \gamma_2=\varphi_{,12} \ .
\label{eq:kappagamma}
\ee
$\Sigma$ is the projected mass density and $\Sigma_{\rm crit.}=(c^2$ /
$4\pi G$)($D_{os}/ (D_{ol} D_{ls})$) is the
critical mass density for which  a light beam emitted    
by a source would exactly focus on the observer plane.\\
The total magnification is  $\mu=1 /((1-\kappa)^2- \vert \gamma^2 \vert )$. Strong magnification events (arcs, microlensing) correspond to cases
where $\mu$ diverges to infinity. At these positions, image multiplications
and strong magnification/distortion occur. For simple lens configurations, the
   location of the infinite amplification in the image plane 
can be easily computed. They
are called  critical lines and  the corresponding positions in the
source planes are the caustic lines. Despite the transient nature
of some microlensing events, giant arcs and microlensing
events are the same physical phenomenon. They have     
different applications because of the different scales involved in each
case. Microlensing is perfectly adapted to local events of stars,
whereas arc(let)s are expected in galaxies or clusters. The weak
lensing regime corresponds to a natural extension of arc(let)s where 
$\kappa$ and $\gamma$ become smaller than $1$. The weak lensing regime is 
observed 
 on  extended sources and is well suited to study  large-scale structures 
 from the correlated alignement of background galaxies. In
fact, since the lensing equation relates mass distribution and cosmological
distances, gravitational lensing can be used in various contexts and
has a large number of applications, as it is shown in figure
\ref{application}.
\section{Microlensing and the baryonic content of the spiral galaxies}\label{sec:baryongal}
For a single point-mass deflector with mass $M_{lens}$, the critical line is a 
circle whose radius is called the Einstein radius and writes
\be
R_e=\sqrt{{4 G M_{lens} \over c^2} {D_{ol} D_{ls} \over D_{os}}} \ .
\label{eq:microre}
\ee
A light beam approaching the lens with an impact parameter $b$ form two
images, separated by 
\be
\theta_{1,2}=\sqrt{b^2+4 R_e^2}=R_e \sqrt{1+{4 \over u^2}} \ ,
\label{eq:microtheta12}
\ee
where $u=b/R_e$. The total amplification of the two images writes
\be 
A ={u^2+2 \over u \sqrt{u^2+4}} \ .
\label{eq:microa}
\ee
Microlensing eventually occurs when $\theta_{1,2}$ is smaller than the
angular resolution of telescopes. For the observer, the images are merged
and only a strong point-like amplification is observed. As an example,
 for the EROS1/2 or MACHO experiments on Magellanic Clouds, the angular
separation  of images is about 1 milliarcsecond which is far below the
best resolutions obtained with ground based telescopes or the HST. \\
In practice, microlensing events are extremely rare. However,
 Paczy\'nski \cite{pac} pointed out that if galactic halos are formed with
compact objects, they move within the galactic potential well  and
eventually can cross light beams of background stars. Transient
microlensing events may thus occur and  
  can be observed if  millions of stars are followed up simultaneously during a
long period.\\
For compact lenses moving with velocity $v(t)$ 
the magnification event 
 spreads over a time scale $\Delta t= R_e/v(t)$. Hence, the typical
time-scale of microlensing events is proportional to $\sqrt{M_{lens}}$ 
 and this property  can be used to probe the mass distribution of compact 
deflectors.
\\
Paczy\'nski's suggestion inspired numerous new projects with the aim
to provide on a short time scale the baryon fraction of our Galaxy
and possibly of extragalactic systems. Table \ref{tablemicrolens}
summarizes the present-day results. We only report single events, so 
 binary-lenses as the one discovered by Alard et al. \cite{amg} are   
not included. The last EROS2 result is the most recent of the table 
 \cite{palanque}.
\begin{table}
\begin{center}
\begin{tabular}{|p{1.2truecm}|p{2.5truecm}|c|p{1.truecm}|p{2.5truecm}|} \hline
{\bf Survey } &{\bf Target} & {\bf Origin} &  
  {\bf Nb of Events} & {\bf Mass Range of the Lens} \\
\hline
EROS1& LMC/SMC & France & 2 &  $\approx 0.15$ M$_{\odot}$ \\
MACHO& LMC/SMC & USA/Australia & 6 &  $\approx 0.3-0.5$ M$_{\odot}$  \\
& Galaxy (bulge)& USA/Australia & $\approx 100$ &  $0.08-0.6$
M$_{\odot}$  \\
OGLE& Galaxy (bulge)  & USA/Poland & 18 &  $0.08-0.6$ M$_{\odot}$  \\
DUO& Galaxy (bulge)  & France   & 13 & $0.08-0.6$ M$_{\odot}$   \\
AGAPE& M31 & France & 0  &  --  \\
KPNO& M31 & USA & 6  &  $\approx 1.0$  M$_{\odot}$ \\
EROS2& LMC/SMC & France & 1 & $0.85-8.7$ M$_{\odot}$   \\
& Galaxy (bulge) & France   & -- & --  \\
\hline
\end{tabular}
\end{center}
\caption{\label{tablemicrolens}The microlensing surveys. 
 AGAPE and the
KPNO/VATT experiments use pixel monitoring since stars cannot be
resolved. AGAPE has not finished yet the data processing. 
A program on M87 with the HST is in project.}
\end{table}

The conclusions of EROS1 and MACHO monitorings are similar and
rule out the possibility that the halo content of our Galaxy has more that
20\% of compact objects with masses ranging between $\approx$ 10$^{-7}$
M$_{\odot}$ and 0.02 M$_{\odot}$. On the other hand, the statistical 
distribution of the events detected in the bulge of the Galaxy is 
compatible with a Salpeter mass distribution with masses ranging between
$0.08$ M$_{\odot}$ and $0.6$ M$_{\odot}$. 
 Complementary results by Crotts \& Tomaney \cite{ct} in M31 
seem to rule out also compact objects with mass between 0.003 and 0.08  
M$_{\odot}$ in 
the outer bulge and the inner disk of this galaxy, whereas their 6 events 
have typical signatures expected for $\approx 1.0$  M$_{\odot}$ objects.
\\
The present-day results of the microlensing experiments are puzzling and
address new  questions on the nature of the low-mass objects
in the Galaxy. The average mass of the objects detected in the direction 
 of the LMC and SMC is 0.5 M$_{\odot}$. Assuming a simple mass
distribution for the halo it is then possible to infer their projected
number-density and to predict that a (small) amount of such objects 
should be detected in the Hubble Deep Field. However, negative
results have been reported so far \cite{fgb}. Similarly,
although the mass function of objects in the direction of the bulge
looks like a Salpeter mass distribution with masses ranging between
$0.08$ M$_{\odot}$ and $0.6$ M$_{\odot}$, the deepest star counts impose 
 strong limits on their number-density and cannot
explain more than 50\% of the  events. In both cases,  a 
significant fraction of the
objects responsible for microlensing events seems somewhat unusual. \\
It is possible that the compact lenses are red-dwarfs,
 but they cannot  account for more than   50\% of the halo. White-dwarfs 
seem unprobable because their progenitors 
 have not been detected, though they 
 should be visible at high-redshift on the deepest 
 astronomical observations \cite{charlotsilk}.
But there are now crucial issues on the existence of a 
large amount of brown dwarfs in our halo in order to explain the flat
rotation curve of our Galaxy. So, even if new fascinating objects are
discovered from the microlensing experiments, they do not 
 provide evidences that  huge baryonic halos of dark matter exist around 
spiral 
galaxies. On the other hand, it seems impossible to abandon the idea
that galaxies do have dark halos. Specific cases of Einstein-ring  
configurations by galaxies \cite{kochi} and recent galaxy-galaxy lensing analyses
 by Brainerd et al. \cite{bbs} and Griffiths et al. \cite{gcr}
 indicate that mass-to-light 
ratios of galaxies
are larger than 10 and that their halos could 
extend up to $\approx$ 100 $h^{-1}$kpc.  
\section{Arc(let)s in clusters of galaxies}\label{sec:arclusters}
Giant arcs form at the points where $\mu$ is (close to) infinite. 
From the position of the source with respect to the caustic lines, one
can explain typical  lensing configurations where arcs form    
from the merging of two, three, or even more images, as well
 as more surprising shapes like  radial arcs and
``straight arcs'' \cite{fm}$^{\!,\,}$ \cite{nb}.\\
Typical lensing cases have been observed in rich clusters
of galaxies, like in MS2137-23 \cite{mfk}, A370
 \cite{kmfm}, AC114 \cite{sces}, or A2218 \cite{kmpmelbbp}.  In some
cases  the positions and shapes of images have been predicted     
and confirmed later which gives strong confidence in the reconstructed
mass distribution. But the most impressive results were provided by
using the HST images (see figure \ref{panelarc}). The outstanding 
image quality of
the HST reveals small details in each arc that can be used
to recognize the multiply-imaged galaxies by eye 
 \cite{wkk}$^{\!,\,}$ \cite{kescs}$^{\!,\,}$ \cite{htsgl}$^{\!,\,}$ \cite{nkse}$^{\!,\,}$ \cite{ssbhbz}.

Arcs constrain the central mass distribution 
 of clusters of galaxies on a scale of $\approx 500$\hkpc
(see table \ref{tablearc}). They have demonstrated that
 their  mass-to-light ratio  on this scale ranges between 100 and 300 
 which definitely shows that 
clusters of galaxies are dominated by dark matter and that $\Omega$
inferred from arcs lies in the range 0.15 to 0.3.
  The lens modelling shows that on  $\approx 100$ kpc scale, 
 dark matter  closely follows the
geometry of the  light distribution associated with the brightest
cluster members. Furthermore, since arcs
occur when $\Sigma /\Sigma_{\rm crit.}>1$, clusters of galaxies must
be much more concentrated
than it was expected from their galaxy and X-ray gas
distributions. But the occurrence of arcs is also enhanced by the
existence of additional clumps observed in most of the rich clusters which
 increases the shear substantially \cite{bar}. The direct
observations of substructures by using  HST images of lensing-clusters
 confirm that clusters are dynamically young systems. 
 Most of these clumps are centered around bright
early-type galaxies (see figure \ref{panelarc}). 
Natarajan \& Kneib \cite{nk} and Geiger \& Schneider \cite{gs} 
 proposed to use these images in order to infer 
the mass of halos of galaxies in clusters. The benefit of additional 
 convergence  by the extended halo of the lensing-cluster enhances the
galaxy-galaxy lensing effect which thus can be easily detected. Though
promising, no conclusive results have been raised yet, mainly because the
two lensing effects (cluster and galaxy-galaxy) mix together and 
combine  weak and strong lensing effect simultaneously 
which are  difficult to separate.

\begin{figure}
\caption{\label{panelarc}HST images of arcs: top left is A2218, center
left: A2390, bottom left: MS2137-23, top right: A370, middle right:
Cl2244-02, bottom right: Cl0024+17.}
\end{figure}
\begin{table}
\begin{center}
\begin{tabular}{|l|c|c|c|c|} \hline
{\bf Cluster } & {\bf z$_{cluster}$} & {\bf Scale} & {\bf M/L} &
{\bf  z$_{arc}$}  \\
 & & ($h^{-1}$~kpc) & ($h$) &  \\
\hline
MS2137-23 \cite{mfk}& 0.33 & 500 & 340-140  &0.7-1.5   \\
\hline
A370 \cite{kmfm}& 0.37 & 500 & $>$ 150  &0.725    
\\
\hline
Cl0024+17 \cite{kkf}& 0.39 & 500 & $>200$  & $\approx$ 0.9   
 \\
\hline
Cl0024+17 \cite{wkk}& 0.39 & 500 & $>200$ &$\approx$ 0.9  
 \\
\hline
A2218 \cite{kmpmelbbp}$^{\!,\,}$ \cite{kescs}& 0.18 & 1000  & 200  &0.702 
\& 1.034   
   \\
\hline
A2390 \cite{plbsk}& 0.23 & 400 & 250  &0.913     
  \\
\hline
\end{tabular}
\end{center}
\caption{\label{tablearc}Mass model inferred from modelling of 
 some spectacular arcs. The scales are expressed in
$h_{100}^{-1}$ Kpc.}
\end{table}

\section{Weak lensing by clusters of galaxies}\label{sec:wlcluster}

Far from the critical lines, light beams of background galaxies crossing 
lensing-clusters are only weakly magnified, but still produce a small 
increase of their
ellipticity in the direction perpendicular to the gradient of the
projected potential (shear). The effect on each individual galaxy is 
much weaker than the average intrinsic  ellipticity of the sources
but it is possible to
use this coherent polarization pattern statistically in order 
to recover the mass distribution of the lens. \\
The population of weakly distorted galaxies probes the distortion
 induced by the lens on  the projected space.  The projected mass
density $\Sigma$ can be recovered by using the  
  shape parameters of the images $M^I$ which are related to the shape 
parameters of the sources $M^S$ by the equation,

\be
M^{S}= A M^I  A,
\label{eq:matricestoi}
\ee
where $M$ are the second  moment of galaxies and  $A$
the magnification matrix. The ellipticity of the sources
$\epsilon_S=(a_S-b_S)/(a_S+b_S)$, where $a_S$ and $b_S$
are the major and minor axis respectively,  
is related to ellipticity of the images
$ \epsilon_I$ and the complex shear $g$ as follows:
\be
\epsilon_S=
{ \epsilon_I- g \over 1- g^{*}  \epsilon_I} \ ;
\  g={ \gamma\over 1-\kappa}\ .
\label{eq:epsilons}
\ee
The sources are randomly distributed, hence their
averaged intrinsic ellipticity is
$\mg \epsilon_S\md=0$ which implies that 
\be
 g=\mg \epsilon_I\md\ .
\label{eq:coeffg}
\ee
In the weak lensing regime,  $(\kappa, \gamma) \ll 1$, and the relations between the physical ($\gamma$) and
observable ($\epsilon_I$) quantities reduce to,
\be
\mg\epsilon_I\md= \gamma\ .
\label{eq:epsiloni}
\ee
The projected mass density $\Sigma$ of the lens is recovered from the
distortion field by using Eq.(\ref{eq:epsiloni}) and the integration of 
Eq.(\ref{eq:kappagamma})
 \cite{ks}$^{\!,\,}$ \cite{ss1}$^{\!,\,}$ \cite{s}$^{\!,\,}$ \cite{ss2}:
\be
\kappa(\vec \theta_I)={-2\over \pi} \int \d^2{ \theta}\
 \ \ \gamma(\vec \theta_I)  \ {\chi(\vec
\theta-\vec \theta_I) \over (\vec \theta-\vec
\theta_I)^2} +\kappa_0 \ ; \ \ \ 
\chi(\vec \theta)=({\theta_1^2- \theta_2^2 \over \theta^2} \ + \ i \ \  {2\theta_1 \theta_2 \over \theta^2}),
\label{eq:kappaetksi}
\ee
where $\kappa_0$ is an integration constant. 

Table \ref{tablearclet}  summarizes results on clusters for which mass
reconstruction from lensing
inversion have been done. When compared with mass distribution
inferred from strong lensing, there is a clear trend towards
 an increase of $M/L$ with  scale.  Bonnet et al. \cite{bmf} find $M/L$ close to
600 at 2.5 \hMpc
from the cluster center, and on 1-\hMpc scale $M/L$ ranges between 
150 and 400, that is $0.2<\Omega<0.5$. This large spread 
 is due to important uncertainties. In particular, the redshift
distribution of the background sources is poorly known. 
Though it  is not a critical issue for
nearby clusters ($z_l<0.2$), because then ${D_{os}/ D_{ls}}\simeq Cste$,
 it could lead to large mass uncertainties for more distant clusters, as
for  MS1054 which is at $z=0.83$ \cite{lk}. The other critical point 
is the measurement of ellipticities as low as 1\% on faint objects 
which is extremely difficult in practice.

Even if the redshift of the sources were known, it is not 
sufficient to get
the absolute value of the mass distribution, because 
 structures with constant mass density eventually intercepting the beam 
do not change the shear map.  This issue is expressed by 
the unknown integration constant $\kappa_0$ in Eq.(\ref{eq:kappaetksi}).
The degeneracy may be broken if one
measures the magnification $\mu$, by comparing properties of lensed
and unlensed sources.  Broadhurst et al. \cite{btp} proposed to
compare the number count
$N(m,z)$ and/or $N(m)$ in a lensed and an unlensed field to measure
$\mu$ (see Fort, this conference). 
 This method was applied on
the cluster by Broadhurst \cite{broad}in A1689 and by Fort et al. \cite{fmd} on
Cl0024+17 and A370. The magnification may also be determined
by the changes of the image sizes at fixed surface brightness \cite{bn}.
 However, even if the measurement of the magnification is an exiting
 approach to probe the lensing effects indepently of the measure
of the shear, it is still extremely difficult to get its amplitude
accuratly. The number counts method is
sensible to the Poisson noise and the intrinsic clustering. The surface
brightness method suffers
of the lack of precise operational definition and of the
dilution of light due to the seeing.

\begin{table}
\begin{center}
\begin{tabular}{|l|c|c|c|c|} \hline
{\bf Cluster } & {\bf z$_{cluster}$} & {\bf Scale} & {\bf M/L} &
{\bf  z$_{source}$}  \\
 &  & ($h^{-1}$~kpc) & ($h$) &   \\
\hline
1455+22 \cite{sef}& 0.26 & 500 & 460  & --
 \\
\hline
Cl0016+16 \cite{sef}& 0.55 & 500 & 430& --
  \\
\hline
MS1224 \cite{fksw}& 0.33 & 500 & 800  & 1-2
  \\
\hline
Cl0024 \cite{bmf}& 0.39 & 2500 & 600  & 1-2
  \\
\hline
A1689 \cite{tf}& 0.18 & 1000 & $400$ &0.9
 \\
\hline
A1689 \cite{k3}& 0.18 & 500 & $>200$ &1-2
 \\
\hline
A1689 \cite{broad}& 0.18 & 500 & $>200$ & 1-2
 \\
\hline
A2218 \cite{skbfwnb}& 0.18 & 400 & 440  &0.7 \&1-2 
  \\
\hline
A2390 \cite{skfbw}& 0.23 & 1000 & 320  &1-2
 \\
\hline
Cl0939 \cite{skss}& 0.41 & 400 & 200  &0.6-1
\\
\hline
MS1054 \cite{lk}& 0.83 & 400 & 1600  &$<$1
  \\
  & & & 580&=1.5 \\
  & & &350 &=3. \\
\hline
\end{tabular}
\end{center}
\caption{\label{tablearclet}Example of mass reconstruction 
obtained from weak lensing inversion. The mass-to-light ratios are higher than 
 from giant arc (see table \ref{tablearc}). The 
uncertainty in the case of MS1054 expresses the dependence
of mass with the redshift of the sources.}
\end{table}

\section{The matter distribution on very large scale}\label{sec:wllss}

The direct observation of the mass distribution on scales larger than
10 \hMpc (or $\approx $ 1 square degree) is a natural step beyond
clusters of galaxies and is actually the main 
 goal in cosmology  for the next decade.  Two observational directions
are now being investigated. In the first one,  
the statistical properties of weakly lensed background galaxies on
degree scales are investigated in order to build maps of the projected 
mass-density of large-scale structures from the gravitational shear, and to
 constrain the cosmological parameters and the projected 
  power spectrum of initial fluctuations.
In the second one, the gravitational shear on small scales ($<10$
arcminutes)  induced 
by non-linear systems is analyzed in order to constrain 
 the cosmological parameters and the projected power spectrum as well,
 and to correct quasar statistics from magnification biases.

At very large scale,  
the lenses are not individually identified, but viewed as
a random population affecting the shape of the galaxies. 
 Each object along a given line of sight 
 may play simultaneously the role of a lens 
for background galaxies and of a source for foreground systems, so that
 the efficiency  depends on the redshift distribution of the 
 whole population.  The shear measured of each field  
 is filtered at a given angular scale so that a signal of
cosmological interest can be extracted.
For a filtering scale of about one degree, the structures
responsible of the gravitational shear being at a redshift
of about 0.4 are expected to be
on scales above 10 \hMpc, that is in a regime where their properties
can be easily predicted with the linear or perturbation theory.
 Blandford et al. \cite{bsbv}, Miralda-Escud\'e \cite{me} and Kaiser \cite{k1}
 argued that the projected power spectrum
should be measurable with such a method
provided shape parameters are averaged on the degree scale, as
it is illustrated on figure \ref{figludo} \cite{mvwbf}. Further studies using various
cosmologies, various statistics and more realistic ellipticity and
redshift distributions have been done recently in the perspective 
of wide fields surveys \cite{villum}$^{\!,\,}$ \cite{bwm}$^{\!,\,}$ \cite{steb}$^{\!,\,}$ \cite{k2}.

\begin{figure}
\caption{\label{figludo}
Simulation of shear induced by large-scale structures. The left
panel shows a (256 Mpc)$^3$ box indicating the
location of matter. The filaments are large-scale structures
which grew from a uniform distribution and an initial power
spectrum $P(k)=k^{-1}$, evolving under the adhesion approximation.  The right
panel shows  thin straight lines which illustrate the
local orientation and intensity of the shear. 
}
\end{figure}

Using the Perturbation Theory,  Bernardeau et al. \cite{bwm} have
analyzed the statistical properties of the gravitational shear averaged
on degree scale. Assuming a power-law for the power spectrum, they 
have shown that  the observation of
about 25 such fields 
 permits to recover the projected power spectrum and $\Omega$
independently, with 10\% accuracy, 
by using the variance,$\mg\kappa^2_{\theta}\md$,
 and the skewness, $\mg\kappa^3_{\theta}\md / 
\mg\kappa^2_{\theta}\md^2$,   
of the probability distribution function
of the local convergence in the sample.
These moments write,
\be
\mg\kappa^2_{\theta}\md \propto \ \ P(k)\ \Omega^{1.5} \ \ z_s^{1.5} \
\ ;
\ \ 
{\mg\kappa^3_{\theta}\md \over \mg\kappa^2_{\theta}\md^2} \propto
\ \ \Omega^{-0.8} \ \ z_s^{-1.35}
\label{eq:varianceetskewness}
\ee
where $\theta =30'$ is the scale where the convergence $\kappa$ is
averaged, $P(k)$ is the projected power spectrum of the dark matter
and $z_s$ is the
averaged redshift of sources. Bernardeau et al. notice 
that the product of these two quantities
 provides an information on $\Omega$ and $P(k)$ regardless 
the redshift distribution of sources which is basically unknown for the
faint galaxies.

Recently, Jain \& Seljak \cite{js} have investigated the effect of
 the non-linear evolution of the power spectrum on the amplitude of the
variance. They show that at scales above 30 arcminutes the variance is
not much affected, but at scales below 2 arcminutes the non-linear
structures induce a significant increase of the expected 
 signal for the shear. Its amplitude is expected to be 
 between 2\% and 5\% depending on cosmology.  
In fact, this ``cosmic shear'' may have been detected
already by Fort et al. \cite{fmdbk} and Schneider et al. \cite{svwmjsf}. If it is so,
we could be already able to obtain a preliminary value for $\Omega$ 
 very soon by using weak lensing effects.

\section{Conclusions}
Microlensing surveys have shown that the existence of a baryonic halo
dominated by brown dwarfs around our Galaxy is not confirmed.   However,
 galaxy-galaxy lensing and Einstein rings suggest that galactic halos 
of dark matter are present. Instead, microlensing events reveal a
new population of compact star-like objects may exist in our Galaxy. On
the other hand, studies of arc(let)s and weak lensing in clusters show
that $\Omega>0.2$ is almost certain and provide evidence that 
 $0.2<\Omega<0.6$ on scales below 2.5 $h^{-1}$ Mpc. The next decade will
 provide important constraints on $\Omega$ and  $P(k)$, as well
as first  maps of mass-density of the Universe on 100$h^{-1}$ Mpc scales
regardless the light distribution.   This is an enthusiastic 
period for gravitational lensing applications to cosmology.

\section*{Acknowledgments}
We thank B. Fort,  P. Schneider and S. Seitz for  fruitful discussions 
and  enthusiastic  collaborations.  We thank B. Fort, 
B. Geiger and J.-P. Kneib
for careful reading of the manuscript.

\end{document}